\documentclass[a4paper, draft,12pt]{article}
\usepackage{amsmath,amssymb}
\begin{document}
\title{\textbf{Representations of quantum superalgebra $U_q[gl(2|1 )]$ in a
coherent state basis and  generalization}}
\author{
{\bf Nguyen Cong Kien$^{1,2}$, Nguyen  Anh Ky$^{1,2}$, Le Ba Nam}$^{3,}$\footnote
{Present address: Department of Physics, University of South Florida,
4202 E. Fowler Avenue, Tampa, FL 33620, USA.} ~ and\\ {\bf Nguyen Thi Hong Van}$^1$\\[2mm]
$^1$Institute of Physics, Hanoi, Vietnam\\
$^2$College of Science, Vietnam National University, Hanoi, Vietnam\\
$^3$Hanoi University of Technology, Hanoi, Vietnam}
%\date{\today}
\maketitle
\begin{abstract}
The coherent state (CS) method has proved to be useful in quantum physics and mathematics. This method, more precisely, the vector coherent state (VCS) method, has been used by some authors to construct representations of superalgebras but almost, to our knowledge, it has not yet been extended to quantum superalgebras, except $U_q[osp(1|2)]$, one of the smallest quantum superalgebras. In this article the method is applied to a bigger quantum superalgebra, namely $U_q[gl(2|1)]$, in constructing $q$--boson-fermion realizations and finite-dimensional representations which, when irreducible, are classified into typical and nontypical representations. This construction leads to a more general class of $q$--boson-fermion realizations and finite-dimensional representations of $U_q[gl(2|1)]$ and, thus,
at $q=1$, of $gl(2|1)$. Both $gl(2|1)$ and $U_q[gl(2|1)]$ have found different physics applications, therefore, it is meaningful to construct their representations.
\end{abstract}
{\it Running title:} Representations of $U_q[gl(2|1)]$ in a coherent state basis
\section{Introduction}

Symmetry groups (and their infinitesimal version -- algebras), including also super-groups and quantum (super) groups (QG's), and their representations play an extremely important role in the development of modern physics and mathematics. In many cases, mathematical needs and physics applications require explicit constructions of representations of groups or  algebras. Depending on motivations and problems considered, different methods of building representations of a group or an algebra can be chosen, and the boson-fermion realization method is one of them. The first type of boson-fermion realizations, or the so-called Jordan-Schwinger map (a boson realization), was suggested by P. Jordan \cite{jordan} and J. Schwinger \cite{Schw65}. Other types were introduced by F. Dyson \cite{Dy56} or by T. Holstein and H. Primakoff \cite{HP40}. The Dyson's method was later generalized for Lie superalgebras \cite{AL92,Pal97} as well as for quantum algebras \cite{BN90} -- \cite{palev98}. The Jordan-Shwinger and Holstein-Primakoff realizations of quantum algebras were also made in \cite{Mac89,Bie89} and \cite{CEK90,Pro93}, respectively. To build a group or algebra representation via a boson-fermion realization, it is convenient to use the coherent state method which has proved to be very efficient in investigating dynamic symmetries (mathematically described by groups/algebras) of quantum systems \cite{malkin}.\\

The concept of coherent states (CS's) was first introduced by E. Schr\"odinger over 80 years ago \cite{Schr26}, and much later, in 1972, it was generalized for an arbitrary Lie group by A. Perelomov (see \cite{Pere72} or more \cite{Pere86,Klau85} and their references). Further, this concept was extended to that of the so-called vector coherent states (VCS's) \cite{vcs1} -- \cite{vcs6} and also to that of super-coherent states (SCS's)  \cite{scs1} -- \cite{scs7}. Hereafter the terminology CS's also contains VCS's and SCS's. The CS's play an essential role in physics and mathematics such as quantum optics \cite{Mandel}, many-body physics \cite{Klau85}, quantum gravity and cosmology \cite{cs-qg, cs-qc}, string theories and integrability \cite{cs-st, st-sm}, path integrals \cite{Fujii1} -- \cite{sun-cs}, quantum information (see \cite{Fujii4} and references therein), representation theory of Lie (super) groups and (super) algebras \cite{Pere72,Pere86,scs5,cs-indrep}, non-commutative geometry \cite{cs-ncg}, etc. A natural relation between CS's and QG's (including quantum supergroups) was also established \cite{Mac89,Bie89,CEK90,Jurco, Cele1,Cele2}. As is well known \cite{vcs1, vcs2, vcs5,scs5,Fujii2,cs-indrep}, the CS theory is tightly related to the theory of induced representations of (super) groups and (super) algebras and also of those of QG's. However, sometimes the CS method for construction of irreducible representations is simpler than the ordinary induced representation methods (Chevalley-Harish-Chandra and Kac methods) and has more direct physics applications. Therefore, the CS representation method sometimes is preferable to the ordinary induced representation methods.\\

Using the CS method (for a more general construction, see, for example, \cite{vcs5,scs5,scs6}) Y. Zhang et al \cite{ZG04,Zh04} constructed representations of some superalgebras such as $osp(1|2)$, $osp(2|2)$ and $gl(2|2)$. However, to our knowledge, this method almost has not yet been used for constructing representations of quantum superalgebras, except $U_q[osp(1|2)]$, one of the smallest quantum superalgebras \cite{Zh04}. In this paper (Sect. 2), we deal with $U_q[gl(2|1)]$, a bigger quantum super algebra. As a result, representations ($q$--boson-fermion realizations and the corresponding finite-dimensional representations) of $U_q[gl(2|1)]$ in a CS basis and their generalization are obtained. Putting $q=1$ we get generalized representations of $gl(2|1)$.
Both $gl(2|1)$ and its quantum deformation, $U_q[gl(2|1)]$, are important for physics applications (see, for example, \cite{susy-e} -- \cite{Tsuboi} and references therein), therefore, to construct their representations is meaningful. Our conclusion is made in the final section.
\section{Representations of $U_q [gl(2|1)]$ in a CS basis}
\vspace*{2mm}
In this section, we will construct representations ($q$--boson-fermion realizations and finite-dimensional representations) of $U_q [gl(2|1)]$ in a coherent state basis
(for other representations of $U_q [gl(2|1)]$ see, for example, \cite{KV03} and references therein). Firstly, we recall the definition of the quantum super algebra $U_q [gl(2|1)]$.
\subsection{Quantum superalgebra $U_q [gl(2|1)]$}
\vspace*{2mm}

The quantum super algebra $U_q [gl(2|1)]$ as a deformation of the universal enveloping algebra  $U[gl(2|1)]$ of the superalgebra $gl(2|1)$ can be defined through its Weyl-Chevalley generators $E_{12} ,\,\,E_{21} ,\,E_{23} ,\,E_{32} $ and $E_{ii}$ $(i = 1,2,3)$ satisfying the following defining relations (see, for example, \cite{KV03}): \\

a) \emph{the (anti-)commutation relations} ($1\leq i,j,i+1,j+1\leq 3$):
\begin{subequations}\label{rlq}
\begin{equation}
[E_{ii} ,E_{jj} ] = 0,
\end{equation}
\begin{equation}
[E_{ii} ,E_{j,j + 1} ] = (\delta _{ij}  - \delta _{i,j + 1} )E_{j,j + 1},
\end{equation}
\begin{equation}
[E_{ii} ,E_{j + 1,j} ] = (\delta _{i,j + 1}  - \delta _{i,j} )E_{j + 1,j},
\end{equation}
\begin{equation}
[E_{12} ,E_{21} ] = [H_1 ]_q,
\end{equation}
\begin{equation}
\{ E_{23} ,E_{32} \}  = [H_2 ]_q,
\end{equation}
\begin{equation}
H_i  = (E_{ii}  - \frac{{d_{i + 1} }}{{d_i }}E_{i + 1,i + 1} ),~~ H_3\equiv E_{33},
\end{equation}
\end{subequations}
where $d_1=d_2=-d_3=1$, while
\begin{equation}\label{E:eq1}
[X]_q := \frac{{q^X - q^{-X} }}{{q - q^{ - 1} }},
\end{equation}
is a quantum deformation ($q$-deformation, for short) of an operator or a number $X$, and\\

b) \emph{ the Serre relations:}
\begin{subequations}\label{Serre}
\begin{equation}
E_{23}^2  = E_{32}^2  = 0,
\end{equation}
\begin{equation}
[E_{12} ,E_{13} ]_q  = [E_{21} ,E_{31} ]_q  = 0,
\end{equation}
\end{subequations}
with $E_{13}$ and $E_{31}$,
\begin{equation}\label{sub}
E_{13}  = [E_{12} ,E_{23} ]_{q^{ - 1} },~~~E_{31}  =  - [E_{21} ,E_{32} ]_{q^{ - 1} }.
\end{equation}
defined as new (non-Chevalley) generators, where $[A,B]_r:=AB-rBA$ is a deformed commutator.
The generators $E_{ij}$, $i,j=1,2,3$, are $q$-deformations ($q$-analogs) of the Weyl generators
$e_{ij}$ of $gl(2|1)$ subject to the supercommutation relations
 \begin{equation}
\left[e_{ij} ,e_{kl}\right\} = \delta _{jk} e_{il}  - ( - 1)^{([i] + [j])([k] + [l])} \delta _{il} e_{kj},
\end{equation}
with  $\left[e_{ij} ,e_{kl} \right\}\equiv e_{ij} e_{kl}  - ( - 1)^{([i] + [j])([k] + [l])} e_{kl} e_{ij}$
being a super-commutator (commutator or anti-commutator) and $[i]$ being a parity defined by $[1]=[2]=0$,
 $[3]=1$.\\

  The subspace of $U_q [gl(2|1)]$ developed on the Cartan generators
 $H_i$ and the even Chevalley generators $E_{12}$ and $E_{21}$ is its even subalgebra which denoted by
 $U_q [gl(2|1)_{\bar{0}}]$ is a quantum deformation of $U[gl(2|1)_{\bar{0}}]$. We will work in this paper with generic $q$, i.e., when it is not a root of unity. The case of $q$ being a root of unity will be investigated elsewhere.
 \subsection{q--boson-fermion realization of $U_q [gl(2|1)]$}
\vspace*{2mm}

In this sub-section, we deal with $q$--boson-fermion realizations of $U_q [gl(2|1)]$.\\

Let $\left| J \right\rangle $  be a state of $U_q [gl(2|1)]$ defined by
\begin{gather}
 H_1 \left| J \right\rangle  = 2J_1 \left| J \right\rangle ,\,H_2 \left| J \right\rangle  = 2J_2 \left| J \right\rangle ,
 \,H_3 \left| J \right\rangle  = 2J_3 \left| J  \right\rangle,  \notag\\[2mm]
 E_{12} \left| J \right\rangle  = E_{13} \left| J \right\rangle  = E_{23} \left| J \right\rangle  = 0.
 \label{hw}
\end{gather}
It is called the the highest weight state characterized by a set of three numbers $J_1$, $J_2$ and $J_3$ called the highest weight:
\begin{equation}
\left| J \right\rangle \equiv \left| J_1,J_2,J_3 \right\rangle.
\end{equation}
We can construct a representation of $U_q[gl(2|1)]$ induced from $\left| J \right\rangle $.
When $J_1$ is an integer or a half-integer, but $J_2$ and $J_3$ can be arbitrary complex numbers,
a $U_q [gl(2|1)]$ module, denoted by $W$, induced
from $\left| J \right\rangle$ has a dimension of $8J_1+4$ and contains four sub-spaces, say $V_i$, $i=1,2,3,4$, which are finite-dimensional
modules of the even subalgebra $U_q [gl(2|1)_{\bar{0}}]$ and respectively spanned on the following basis vectors:
\begin{eqnarray}
\notag
%|V^{(1)};M\rangle\equiv
|J_1,J_2,J_3,M\rangle &=&C^{(1)}_M.E_{21}^{J_1-M}|J\rangle\in V_1,\\\nonumber\vspace*{4mm}
%|V^{(2)};P\rangle\equiv
|J_1-1/2,J_2+1/2,J_3+1/2,P\rangle &= & C^{(2)}_P.\left \{q^{J_1+P+1/2}E_{32}E_{21}^{J_1-P+1/2}|J\rangle \right.\\ \nonumber
&&\left. + [J_1+P+1/2]_qE_{31}E_{21}^{J_1-P-1/2}|J\rangle\right \}\in V_2,\\ \nonumber\vspace*{4mm}
%
%
%|V^{(3)};R\rangle\equiv
|J_1+1/2,J_2,J_3+1/2,R\rangle&=&C^{(3)}_R.\left \{\frac{1}{q^{J_1-R+1/2}} E_{32}E_{21}^{J_1-R+1/2}|J\rangle \right.\\ \nonumber
&&\left. -[J_1-R+1/2]_qE_{31}E_{21}^{J_1-R-1/2}|J\rangle\right \}\in V_3,\\ \vspace*{4mm}
%|V^{(4)};S\rangle\equiv
|J_1,J_2+1/2,J_3+1,S\rangle &=& C^{(4)}_S . E_{32}E_{31}E_{21}^{J_1-S}|J\rangle\in V_4,
\label{adbasisgl}
\end{eqnarray}
with $C^{(i)}_K$ being normalization coefficients which can be determined by addtional, for example, Hermitian conditions, and $-J_1\leq M,S \leq J_1$, $-(J_1-1/2) \leq P \leq (J_1-1/2)$,
$-(J_1+1/2) \leq R \leq (J_1+1/2)$ such that $(J_1-M)$, $(J_1-P-1/2)$, $(J_1-R+1/2)$ and $(J_1-S)$ are integers.
These four finite-dimensional irreducible modules of the even subalgebra $U_q[gl(2|1)_{\bar 0}]$ are built on the
$U_q[gl(2|1)_{\bar 0}$]-highest weight states
\begin{eqnarray}
|J_1,J_2,J_3;J_1\rangle &\equiv& |J_1,J_2,J_3\rangle=\left| {J  } \right\rangle,\nonumber \\[2mm]
|J_1-1/2,J_2+1/2,J_3+1/2;J_1-1/2\rangle &\equiv& |J_1-1/2,J_2+1/2,J_3+1/2\rangle,\nonumber \\[2mm]
|J_1+1/2,J_2,J_3+1/2;J_1+1/2\rangle &\equiv & |J_1+1/2,J_2,J_3+1/2\rangle,\nonumber \\[2mm]
|J_1,J_2+1/2,J_3+1;J_1\rangle &\equiv & |J_1,J_2+1/2,J_3+1\rangle,
\end{eqnarray}
via the formula
\begin{eqnarray}
|L_1,L_2,L_3;k\rangle=C^{(i)}_{k}.(E_{21})^{L_1-k}|L_1,L_2,L_3;L_1\rangle, ~~~i=1,2,3,4,
\end{eqnarray}
where $C^{(i)}_{k}$ are normalization coefficients in (\ref{adbasisgl}),
$L_1=J_1,J_1\pm1/2$; $L_2=J_2,J_2+1/2$;
$L_3=J_3,J_3\pm1/2, J_3+1$ and $-L_1\leq k\leq L_1$ such that $L_1-k$ are integers. For simplicity,
we can choose $C^{(i)}_k=1$ and work with vectors (\ref{adbasisgl}) non-normalized.
The odd generators of $U_q[gl(2|1)$ intertwine these
$U_q[gl(2|1)_{\bar 0}]$-modules.\\

Now, generalized coherent states of  $U_q [gl(2|1)]$ can be defined as (cf. \cite{Pere86,Klau85,ZG04}, \cite{scs1}--\cite{scs6})
\begin{equation}
e_q^{a_{12} E_{21}  + \alpha _{23} E_{32}  + \alpha _{13} E_{31} } \left| {J } \right\rangle,
\label{cs}
\end{equation}
where $e_q^X  = \sum\limits_{n = 0}^\infty  {\frac{{X^n }}{{[n]_q!}}} $ denotes the so-called $q$-exponent
with $[n]_q!=[1]_q.[2]_q...[n]_q$ and $[n]_q$ given in \eqref{E:eq1}, while $\alpha_{ij}^\dag  ~(\alpha_{ij} )$ are $q$-analogs of the fermion
creating (annihilating) operators with the number operator $N_{\alpha_{ij} }$
and $a_{12}^\dag ~ (a_{12} )$ is a $q$-analog of the boson creating (annihilating) operator with the number operator $N_{a_{12} }$. They form the quantum Heisenberg
superalgebra $U_q[h(2|1)]$:
\begin{eqnarray}
\label{sqHei21}
\{ \alpha_{ij} ,\alpha_{ij}^\dag  \}  = 1,~~~N_{\alpha_{ij} }  = \alpha_{ij}^\dag  \alpha_{ij} ,~~~[N_{\alpha_{ij} } ,\alpha_{ij}^\dag  ]
= \alpha_{ij}^\dag  ,~~~[N_{\alpha_{ij} } ,\alpha_{ij} ] =  - \alpha_{ij}\nonumber\\
 \left[a_{12} ,a_{12}^\dag\right]_q = q^{ - N_{a_{12} } } ,~~~\left[N_{a_{12} } \right]_q = a_{12}^\dag  a_{12},~~~\left[N_{a_{12} }  + 1\right]_q = a_{12} a_{12}^\dag,~~~~\nonumber\\
 \left[N_{a_{12} } ,a_{12}^\dag \right ] = a_{12}^\dag  ,~~~\left[N_{a_{12} } ,a_{12} \right] =  - a_{12}.~~~~~~~~~~~~~\nonumber
\end{eqnarray}\\
This quantum Heisenberg superalgebra $U_q[h(2|1)]$ super-commutes with $U_q[gl(2|1)]$, that is, if $E$ is an operator of $U_q[gl(2|1)]$ and $X$ is an operator of $U_q[h(2|1)]$ they super-commute with each other:
\begin{equation}
EX=(-1)^{deg(E).deg(X)}XE,
\end{equation}
where $deg(X)$ is the parity of $X$.\\

Let $|\psi\rangle$ be a state vector in a (e.g.,
finite-dimensional) module of $U_q[gl(2|1)]$.
Then the mapping (cf. \cite{ZG04} and note the difference between the notations here and there)
\begin{equation}
\label{eq2}
\left| \psi  \right\rangle  \to \left| \psi  \right\rangle _J=\left\langle {J } \right|e_q^{a_{12}^\dag  E_{12}  + \alpha_{23}^\dag  E_{23}  +
\alpha_{13}^\dag  E_{13} } \left| \psi  \right\rangle \left| 0 \right\rangle,
\end{equation}
induces the mapping
\begin{equation}
\label{eq3}
A \to \Gamma (A)\left| \psi  \right\rangle _J  = \left\langle {J } \right|e_q^{a_{12}^\dag  E_{12}  + \alpha_{23}^\dag  E_{23}+ \alpha_{13}^\dag  E_{13} } A\left| \psi  \right\rangle \left| 0 \right\rangle,
\end{equation}
of an operator $A$ defined in a space, which here is a $U_q[gl(2|1)]$ module, containing $|\psi\rangle$, where $|0\rangle$ is a vacuum state of the quantum Heisenberg superalgebra $U_q[h(2|1)$:
\begin{equation}
a_{12}| 0 \rangle  = \alpha _{ij} | 0 \rangle  = 0.
\end{equation}
 Theoretically, a construction of a finite-dimensional representation of  $U_q[gl(2|1)]$ can be performed step by step as in the case of non-deformed superalgebras. However,  there are underwater rocks preventing a smooth performance of such a construction. One of the obstacles is to expand the expression $e_q^{a_{12}^\dag  E_{12}  + \alpha _{23}^\dag  E_{23}  + \alpha _{13}^\dag  E_{13} } $ into a product of exponents. This problem is not simple as that
for a non-deformed exponent. Here, instead of Baker-Campbell-Hausdorff formula a $q$-deformation of the generalized Zassenhaus formula \cite{q-Zassen} can be used but we prefer a direct computation.
%\begin{equation}
 %e_q^{(X + Y)}  = e_q^X e_q^Y \prod\limits_{i = 2}^\infty  {e_q^{C_i } },
%\end{equation}
%with $C_2  = \frac{1}{2}[X,Y]$, ~$ C_3  = \frac{1}{3}[[Y,X],Y] + \frac{1}{6}[[Y,X],X]$, ~
%$C_4  = \frac{1}{8}([[[Y,X],Y],Y] + [[[Y,X]X],Y]) + \frac{1}{{24}}[[[Y,X],X],X]$, etc.,
%and the nilpotent property of fermion operators.
%However, we prefer to decompose $e_q^{a_{12}^\dag  E_{12}
%+ \alpha _{23}^\dag  E_{23}  + \alpha _{13}^\dag  E_{13} } $
%by a direct computation.
As a result we obtain
%To go further we must first decompose the $q$-exponent
%$e_q^{a_{12}^\dag  E_{12}  + \alpha _{23}^\dag  E_{23}
%+ \alpha _{13}^\dag  E_{13} } $ intro a product of single exponents.
%A a result we obtain
\begin{eqnarray}
 \label{exq}
e_q^{\alpha_{13}^\dag E_{13}+\alpha_{23}^\dag E_{23}+a_{12}^\dag E_{12}}
&=&\left\{D_4(N_{a_{12}})\alpha_{13}^\dag E_{13}\alpha_{23}^\dag E_{23}
+D_3 (N_{a_{12} } )\alpha_{13}^\dag  E_{13} \right. \nonumber \\
&& \left.+e^{D_2(N_{a_{12}})a_{12}^\dag \alpha_{23}^\dag E_{13}}e^{D_1(N_{a_{12}})\alpha_{23}^\dag E_{23}}\right\}e_q^{a_{12}^\dag E_{12}},
\end{eqnarray}
where,
\begin{eqnarray}\label{func}
D_1 (N_{a_{12} } ) &=& \frac{q + 1}{q^{N_{a_{12} }  + 1} + 1}, \nonumber\\[2mm]
D_2 (N_{a_{12} } ) &=& \frac{\left([N_{a_{12} } ]_q  - [N_{a_{12} }  + 1]_q  + 1\right )}{\left(1 - q\right)\left(1 - q^{ - 1} \right)\left[N_{a_{12} }  + 1\right]_q \left[N_{a_{12} } \right]_q }, \nonumber\\[2mm]
D_3 (N_{a_{12} } ) &=& \frac{\left(1 - q^{N_{a_{12} }  + 1} \right)}{\left[N_{a_{12} }  + 1\right]_q \left(1 - q\right)}, \nonumber\\[2mm]
D_4 (N_{a_{12} } ) &= &\frac{(N_{a_{12} }  + 2)(1 - q)\left(1 - q^{ - 1} \right) + \left(1 - q^{N_{a_{12} }  + 2} \right)\left(q^{ - 1}  - q^{ - N_{a_{12} }  - 1} \right)} {\left(1 - q^{ - 1} \right)\left(1 - q\right)\left[N_{a_{12} }  + 2\right]_q\left [N_{a_{12} }  + 1\right]_q }.
  \end{eqnarray}
Using \eqref{rlq}, \eqref{Serre}, \eqref{sub}, \eqref{eq3} and \eqref{exq} we find an explicit boson-fermion realization of the generators:
\begin{subequations}
\begin{equation}
\label{gamma-h1}
\Gamma(H_1)=2J_1 - 2N_{a_{12}}+ N_{a_{23}}-N_{\alpha_{13}}, ~~~~~~~~~~~~~~~~~~~~~~~~~~~~~~~~~~~~~~~~~~
\end{equation}
\begin{equation}
\Gamma (H_2)= 2J_2 + N_{a_{12}}+ N_{\alpha_{13}}, ~~~~~~~~~~~~~~~~~~~~~~~~~~~~~~~~~~~~~~~~~~~~~~~~~~~~
\end{equation}
\begin{equation}
\Gamma (H_3)= 2J_3 + N_{\alpha_{13} }  + N_{\alpha_{23}}, ~~~~~~~~~~~~~~~~~~~~~~~~~~~~~~~~~~~~~~~~~~~~~~~~~~~~
\end{equation}
\begin{eqnarray}
\Gamma(E_{12})&=&\left(a_{12}\alpha_{23}\alpha_{23}^\dag+ a_{12}\frac{{D_1(N_{a_{12}}-1)}} {{D_1(N_{a_{12}})}}N_{\alpha_{23}}\right)\alpha_{13}\alpha_{13}^\dag ~~~~~~~~~~~~~~~~~~~~~~~~~~~~~~~~~~\nonumber\\
&+&  a_{12}\frac{{D_4(N_{a_{12}}- 1)}}{{D_4(N_{a_{12}})}}N_{\alpha_{13}}N_{\alpha_{23}} + \frac{{D_2 (N_{a_{12} } )}}
{{D_3(N_{a_{12}})}}[N_{a_{12}}]_q\alpha_{23}^\dag\alpha_{13}\nonumber\\
&-&  \frac{{D_1(N_{a_{12}})}}{{D_1(N_{a_{12} } )+1}}\frac{{D_2 (N_{a_{12} } )+1}}
{{D_3 (N_{a_{12}})}}[N_{a_{12}}+1]_q\alpha_{23}^\dag \alpha_{13} \nonumber\\
&+&a_{12}\frac{{D_3(N_{a_{12}} - 1)}}{{D_3 (N_{a_{12} } )}}N_{\alpha_{13} } \alpha_{23} \alpha_{23}^\dag ,
%\nonumber\\
\end{eqnarray}
\begin{eqnarray}
\Gamma(E_{21})&=&\left\{\begin{array}{c}
 \frac{D_2(N_{a_{12}})}{D_1(N_{a_{12}}-1)}\left[a_{12}^\dag N_{\alpha_{23}}\alpha_{13}\alpha_{13}^\dag-\frac{D_2(N_{a_{12}}-1)}{D_3(N_{a_{12}- 2)}}\left(a_{12}^\dag\right)^2\alpha_{23}^\dag\alpha_{13}\right] \nonumber\\
  + \frac{D_3(N_{a_{12}})}{D_1(N_{a_{12}})}\alpha_{13}^\dag\alpha_{23}\left(1-\frac{D_2(N_{a_{12}})}{D_3 (N_{a_{12}}-1)}a_{12}^\dag\alpha_{23}^\dag\alpha_{13}\right) \\
 \end{array}\right\} \nonumber\\
 &&\times \left(-q^{-(2J_1 + 1)}\right) \nonumber \\
&+ &\left\{
\begin{array}{l}
\left.[2J_1-N_{a_{12}}+1]_q a_{12}^\dag \alpha_{23}\alpha_{23}^\dag +\right. \\
\left.[2J_1+2-N_{a_{12}}]_q \frac{{D_1(N_{a_{12}})}}{{D_1(N_{a_{12}}-1)}}a_{12}^\dag N_{\alpha_{23}}\right.
\end{array}\right\}\alpha_{13}\alpha_{13}^\dag
\nonumber\\
 &+&\left\{
 \begin{array}{l}
 \left.[2J_1-N_{a_{12}}+1]_q \frac{{D_2(N_{a_{12}})}}{{D_3(N_{a_{12}}- 2)}}-\right.\\
 \left.[2J_1-N_{a_{12}}+ 2]_q \frac{{D_1 (N_{a_{12}})}}{{D_1(N_{a_{12}}-1)}}\frac{{D_2(N_{a_{12}}-1)}}{{D_3(N_{a_{12}}-2)}}\right.
 \end{array}\right\}\left(a_{12}^\dag\right)^2\alpha_{23}^\dag \alpha_{13}  \nonumber\\
 &+& [2J_1- N_{a_{12}}]_q \frac{{D_3 (N_{a_{12}})}}{{D_3(N_{a_{12}}- 1)}}a_{12}^\dag N_{\alpha_{13}}\alpha_{23}\alpha_{23}^\dag \nonumber\\
 &+&[2J_1- N_{a_{12}}+ 1]_q \frac{{D_4(N_{a_{12}})}}{{D_4 (N_{a_{12}}- 1)}}a_{12}^\dag N_{\alpha_{13}} N_{\alpha_{23}} ,
\end{eqnarray}
\begin{equation}
\Gamma(E_{13}) = \frac{{q^{N_{a_{12}}}}}{{D_3(N_{a_{12}})}}\alpha_{13}\alpha_{23}\alpha_{23}^\dag +q^{N_{a_{12}}}\frac{{D_1(N_{a_{12}})}}{{D_4( N_{a_{12}})}}\alpha_{13}N_{\alpha_{23}},~~~~~~~~~~~~~~~~~~~~~~~~~~~~~~~~~
\end{equation}

\begin{eqnarray}
 \Gamma(E_{31})& = &\frac{\left[2J_1+2J_2\right]_q}{q} D_2 (N_{a_{12} } )a_{12}^\dag  \alpha_{23}^\dag  \alpha_{13} \alpha_{13}^\dag \nonumber\\
 &+& \frac{\left[2J_1+2J_2\right]_q}{q}D_3 (N_{a_{12} } )\alpha_{13}^\dag  \alpha_{23} \alpha_{23}^\dag   \nonumber\\
  &+&\frac{\left[2J_1+2J_2\right]_q}{q} \frac{D_4 (N_{a_{12}} )}{D_1 (N_{a_{12}} )}\left[ \alpha_{13}^\dag  N_{\alpha_{23} }  + \frac{{D_2 (N_{a_{12} } )}}{{D_3 (N_{a_{12}} - 1)}}a_{12}^\dag  \alpha_{23}^\dag  N_{\alpha_{13} } \right] \nonumber\\
  &+& q^{2J_2 + 2J_1 } \left[D_4 (N_{a_{12}})\alpha_{13}^\dag  \frac{{N_{\alpha {\kern 1pt} _{23} } }}{{D_1 (N_{a_{12} } )}} + D_4 (N_{a_{12} } )\frac{{N_{\alpha_{13} } }}{{D_1 (N_{a_{12} } )}}D_2 (N_{a_{12} } )a_{12}^\dag  \alpha_{23}^\dag \right] \nonumber\\
 &-& q^{2J_1 + 2J_2 - N_{a_{12}}}[N_{a_{12}} - 1]_q D_2(N_{a_{12}})a_{12}^\dag \alpha_{23}^\dag \alpha_{13} \alpha_{13}^\dag   \nonumber\\
  &-&[2J_2]_q q^{2J_1-N_{a_{12}}+1} D_1 (N_{a_{12}})a_{12}^\dag  \alpha_{23}^\dag \alpha_{13} \alpha_{13}^\dag  \nonumber \\
  &-& q^{2J_1 + 2J_2 - N_{a_{12}} -1} D_3 (N_{a_{12}})[N_{a_{12}}]_q\alpha_{13}^\dag \alpha_{23} \alpha_{23}^\dag   \nonumber\\
  &-& [2J_2 + 1]_q q^{2J_1 - N_{a_{12}}} \frac{{D_4 (N_{a_{12}})}}{{D_3(N_{a_{12}} - 1)}}a_{12}^\dag \alpha_{23}^\dag  N_{\alpha_{13} }  \nonumber\\
  &-& q^{2J_1+2J_2-N_{a_{12}}-2} \frac{D_4 (N_{a_{12}})}{D_1 (N_{a_{12}})}[N_{a_{12}}]_q \alpha_{13}^\dag \left[N_{\alpha_{23}} - \frac{D_2(N_{a_{12}})}{D_3(N_{a_{12}} - 1)}a_{12}^\dag \alpha_{23}^\dag \alpha_{13}\right] \nonumber\\
&+& q^{2J_1 + 2J_2 - N_{a_{12}}} \frac{{D_4 (N_{a_{12}})}}{{D_3(N_{a_{12}}- 1)}}a_{12}^\dag \alpha_{23}^\dag  N_{\alpha_{13} },  %\nonumber\\
 \end{eqnarray}

\begin{eqnarray}
\Gamma(E_{23})&=&\frac{1}{{q^{N_{a_{12}}}D_1(N_{a_{12}})}}\alpha_{23}\alpha_{13}\alpha_{13}^\dag   - \frac{1}{{q^{N_{a_{12}}}D_1(N_{a_{12}})}}\frac{{D_2(N_{a_{12}})}}{{D_3 (N_{a_{12}}
-1)}}a_{12}^\dag \alpha_{13}\alpha_{23}\alpha_{23}^\dag \nonumber \\
&-&  \frac{1}{{q^{N_{a_{12}}- 1}}}\frac{{D_2 (N_{a_{12}})}}{{D_4 (N_{a_{12}}-1)}}a_{12}^\dag \alpha_{13} N_{\alpha_{23}} + \frac{{D_3(N_{a_{12}})}}{{q^{N_{a_{12}}}D_4(N_{a_{12}})}}\alpha_{23}N_{\alpha_{13}} \nonumber\\
&+& \frac{1}{{D_3(N_{a_{12}}-1)}}a_{12}^\dag \alpha_{13}\alpha_{23} \alpha_{23}^\dag + \frac{{D_1 (N_{a_{12}})}}{{D_4 (N_{a_{12}}- 1)}}a_{12}^\dag \alpha_{13} N_{\alpha_{23}}, \end{eqnarray} \nonumber \\
\begin{eqnarray}
\Gamma(E_{32})&=&q^{2J_2}D_2(N_{a_{12}})[N_{a_{12}}]_q\alpha_{23}^\dag \alpha_{13}\alpha_{13}^\dag
+[2J_2 ]_qD_1(N_{a_{12}})\alpha_{23}^\dag \alpha_{13}\alpha_{13}^\dag   \nonumber\\
&+&q^{2J_2}D_3(N_{a_{12}})a_{12}\alpha_{13}^\dag \alpha_{23}\alpha_{23}^\dag + \frac{{D_4(N_{a_{12}})}}{{D_3(N_{a_{12}})}}[2J_2+1]_q \alpha_{23}^\dag N_{\alpha_{13}}\nonumber\\
 &+& q^{2J_2}D_4(N_{a_{12}})\left\{\frac{1}{q}a_{12}\alpha_{13}^\dag  \frac{1}{D_1(N_{a_{12}})}\left[N_{\alpha_{23}}- \frac{{D_2(N_{a_{12}})}}{{D_3(N_{a_{12}} - 1)}}a_{12}^\dag \alpha_{23}^\dag \alpha_{13}\right] \right. \nonumber \\
 && \left. -\frac{{\alpha_{23}^\dag}}{{D_3(N_{a_{12}} )}}N_{\alpha_{13}} \right\}.
\label{gamma-e32}
\end{eqnarray}
\label{Gamq}
\end{subequations}
Applying Maclaurin's expansion for fermion operators
\begin{equation}\label{Mac}
q^{N_{\alpha _{ij} } }  = 1 + qN_{\alpha _{ij} }  - N_{\alpha _{ij} }
\end{equation}
we can prove that the operators \eqref{Gamq} really satisfy the defining relations
(\ref{rlq}) -- (\ref{sub}) of $U_q[gl(2|1)]$. So, the mapping \eqref{eq3} performs a $q$--boson-fermion
realization of $U_q[gl(2|1)]$.

\subsection{Typical and nontypical representations}

For the finite-dimensional representations \eqref{adbasisgl}, the mapping \eqref{eq2} gives
\begin{eqnarray}
|J_1,J_2,J_3,M\rangle_J &=&(a_{12}^{\dag})^{(J_1-M)}|0\rangle, \nonumber \\[2mm]
|J_1-1/2,J_2+1/2,J_3+1/2,P\rangle_J \nonumber
&=&2D_2 (J_1-P+1/2)\alpha _{23}^\dag  \,(a_{12}^\dag  )^{J_1-P+1/2} \left| 0 \right\rangle \nonumber\\
&& + 2D_3 (J_1-P-1/2)\alpha _{13}^\dag (a_{12}^\dag )^{J_1-P-1/2} \left| 0 \right\rangle ,\nonumber\\[2mm]
|J_1+1/2,J_2,J_3+1/2,R\rangle_J \nonumber
&=&2\left\{ \begin{array}{l}
 D_1 (J_1-R+1/2)[2J_1  + 1]_q  \\- D_2 (J_1-R+1/2)[J_1-R+1/2]_q q^{ - 2J_1  - 1}
 \end{array} \right\} \nonumber \\
&&\times \alpha _{23}^\dag (a_{12}^\dag )^{J_1-R+1/2} \left| 0 \right\rangle  \nonumber\\
&&  - 2D_3 (J_1 -R- 1/2)[J_1-R+1/2]_q q^{ - 2J_1  - 1} \nonumber \\
 && \times \alpha _{13}^\dag (a_{12}^\dag )^{J_1-R - 1/2} \left| 0 \right\rangle  ,\nonumber\\[2mm]
|J_1,J_2+1/2,J_3+1,S\rangle_J&=&\alpha_{23}^{\dag}\alpha_{13}^{\dag}(a_{12}^{\dag})^{(J_1-S)}|0\rangle.
\label{map-J}
\end{eqnarray}
Using \eqref{Gamq} we can obtain all the matrix elements of $U_q[gl(2|1)]$ in the basis \eqref{map-J}. The latter constitute subspaces which being $U_q[gl(2|1)_{\bar 0}]$-modules are invariant under the action of even
$U_q[gl(2|1)]$ generators but shifted from one to another by the odd generators:
\begin{eqnarray}
\label{Hi}
&\Gamma(H_1)&|J_1,J_2,J_3,M\rangle_J=2M|J_1,J_2,J_3,M\rangle_J,\nonumber\\[2mm]
&\Gamma(H_1)&|J_1-1/2,J_2+1/2,J_3+1/2,P\rangle_J\nonumber\\
&=&2P|J_1-1/2,J_2+1/2,J_3+1/2,P\rangle_J,\nonumber \\[2mm]
&\Gamma(H_1)&|J_1+1/2,J_2,J_3+1/2,R\rangle_J\nonumber\\
&=&2R|J_1-1/2,J_2,J_3+1/2,R\rangle_J,\nonumber \\[2mm]
&\Gamma(H_1)&|J_1,J_2+1/2,J_3+1,S\rangle_J\nonumber\\
&=&2S|J_1,J_2+1/2,J_3+1,S\rangle_J.
\end{eqnarray}
\begin{eqnarray}
&\Gamma(H_2)&|J_1,J_2,J_3,M\rangle_J\nonumber\\&=&(2J_2+J_1-M)|J_1,J_2,J_3,M\rangle_J,\nonumber\\[2mm]
&\Gamma(H_2)&|J_1-1/2,J_2+1/2,J_3+1/2,P\rangle_J\nonumber\\
&=&(2J_2+J_1-P+1/2)|J_1-1/2,J_2+1/2,J_3+1/2,P\rangle_J,\nonumber\\[2mm]
&\Gamma(H_2)&|J_1+1/2,J_2,J_3+1/2,R\rangle_J\nonumber\\
&=&(2J_2+J_1-R+1/2)|J_1-1/2,J_2,J_3+1/2,R\rangle_J,\nonumber\\[2mm]
&\Gamma(H_2)&|J_1,J_2+1/2,J_3+1,S\rangle_J\nonumber\\
&=&(2J_2+J_1-S+1)|J_1,J_2+1/2,J_3+1,S\rangle_J.
\end{eqnarray}
\begin{eqnarray}
&\Gamma(H_3)&|J_1,J_2,J_3,M\rangle_J\nonumber\\
&=&2J_3|J_1,J_2,J_3,M\rangle_J,\nonumber\\[2mm]
&\Gamma(H_3)&|J_1-1/2,J_2+1/2,J_3+1/2,P\rangle_J\nonumber\\
&=&(2J_3+1)|J_1-1/2,J_2+1/2,J_3+1/2,P\rangle_J,\nonumber\\[2mm]
&\Gamma(H_3)&|J_1+1/2,J_2,J_3+1/2,R\rangle_J\nonumber\\
&=&(2J_3+1)|J_1+1/2,J_2,J_3+1/2,R\rangle_J,\nonumber\\[2mm]
&\Gamma(H_3)&|J_1,J_2+1/2,J_3+1,S\rangle_J\nonumber\\
&=&(2J_3+2)|J_1,J_2+1/2,J_3+1,S\rangle_J.
\end{eqnarray}
\begin{eqnarray}
&\Gamma(E_{12})&|J_1,J_2,J_3,M\rangle_J\nonumber\\
&=&\left[J_1-M\right]_q|J_1,J_2,J_3,M+1\rangle_J,\nonumber\\[2mm]
&\Gamma(E_{12})&|J_1-1/2,J_2+1/2,J_3+1/2,P\rangle_J\nonumber\\
&=&\left[J_1-P-1/2\right]_q|J_1-1/2,J_2+1/2,J_3+1/2,P+1\rangle_J,\nonumber\\[2mm]
&\Gamma(E_{12})&|J_1+1/2,J_2,J_3+1/2,R\rangle_J\nonumber\\
&=&\left[J_1-R+1/2\right]_q|J_1+1/2,J_2,J_3+1/2,R+1\rangle_J,\nonumber\\[2mm]
&\Gamma(E_{12})&|J_1,J_2+1/2,J_3+1,S\rangle_J\nonumber\\
&=&\left[J_1-S\right]_q |J_1,J_2+1/2,J_3+1,S+1\rangle_J.
\end{eqnarray}
\begin{eqnarray}
&\Gamma(E_{21})&|J_1,J_2,J_3,M\rangle_J\nonumber\\
&=&\left[J_1+M\right]_q|J_1,J_2,J_3,M-1\rangle_J,\nonumber\\[2mm]
&\Gamma(E_{21})&|J_1-1/2,J_2+1/2,J_3+1/2,P\rangle_J\nonumber\\
&=&\left[J_1+P-1/2\right]_q|J_1-1/2,J_2+1/2,J_3+1/2,P-1\rangle_J,\nonumber\\[2mm]
&\Gamma(E_{21})&|J_1+1/2,J_2,J_3+1/2,R\rangle_J\nonumber\\
&=&\left[J_1+R+1/2\right]_q|J_1+1/2,J_2,J_3+1/2,R-1\rangle_J,\nonumber\\[2mm]
&\Gamma(E_{21})&|J_1,J_2+1/2,J_3+1,S\rangle_J\nonumber\\
&=&\left[J_1+S\right]_q|J_1,J_2+1/2,J_3+1,S-1\rangle_J.
\end{eqnarray}
\begin{eqnarray}
&\Gamma(E_{13})&|J_1,J_2,J_3,M\rangle_J=0,\nonumber\\[2mm]
&\Gamma(E_{13})&|J_1-1/2,J_2+1/2,J_3+1/2,P\rangle_J\nonumber\\
&=&2q^{J_1-P-1/2}|J_1,J_2,J_3,P+1/2\rangle_J,\nonumber\\[2mm]
&\Gamma(E_{13})&|J_1+1/2,J_2,J_3+1/2,R\rangle_J\nonumber\\
&=&- 2q^{-J_1 - R-3/2} \left[J_1-R+1/2\right]_q |J_1,J_2,J_3,R+1/2\rangle_J,\nonumber\\[2mm]
&\Gamma(E_{13})&|J_1,J_2+1/2,J_3+1,S\rangle_J\nonumber\\
&=&- \frac{\left[ J_1-S \right]_q q^{ - J_1-S  - 1} }{2D_4(J_1-S)\left[ {2J_1  + 1} \right]_q} \nonumber \\
&& \times |J_1-1/2,J_2+1/2,J_3+1/2,S+1/2\rangle_J \nonumber\\
&&-\frac{q^{J_1-S}}{2D_4(J_1-S)\left[2J_1+1\right]_q}|J_1+1/2,J_2,J_3+1/2,S+1/2\rangle_J.
\end{eqnarray}
\begin{eqnarray}
&\Gamma(E_{31})&|J_1,J_2,J_3,M\rangle_J\nonumber\\
&=&\frac{ \left[ {2J_1  + 2J_2 +1} \right]_q \left[J_1 + M \right]_q q^{-(J-M)-1}} {2\left[ 2J_1  + 1 \right]_q } \nonumber\\
 &&\times|J_1-1/2,J_2+1/2,J_3+1/2,M-1/2\rangle_J \nonumber\\
 &&- \frac{\left[ {2J_2 } \right]_qq^{J_1 + M} }{2\left[ 2J_1  + 1 \right]_q}|J_1+1/2,J_2,J_3+1/2,M-1/2\rangle_J, \nonumber \\[2mm]
&\Gamma(E_{31})&|J_1-1/2,J_2+1/2,J_3+1/2,P\rangle_J\nonumber\\
&=&-2q^{J_1 +P-3/2 }[2J_2]_q D_4(J_1-P+1/2)|J_1,J_2+1/2,J_3+1,P-1/2\rangle_J
,\nonumber \\[2mm]
&\Gamma(E_{31})&|J_1+1/2,J_2,J_3+1/2,R\rangle_J\nonumber\\
&=&-2q^{-J_1 +R-3/2}[2J_1+2J_2+1]_q[J_1+R+1/2]_qD_4 (J_1-R-1/2)\nonumber \\
&& \times|J_1,J_2+1/2,J_3+1,R-1/2\rangle_J,\nonumber\\[2mm]
&\Gamma(E_{31})&|J_1,J_2+1/2,J_3+1,S\rangle_J=0.
\end{eqnarray}
\begin{eqnarray}
&\Gamma(E_{23})&|J_1,J_2,J_3,M\rangle_J=0,\nonumber\\[2mm]
&\Gamma(E_{23})&|J_1-1/2,J_2+1/2,J_3+1/2,P\rangle_J\nonumber\\
&=&2|J_1,J_2,J_3,P-1/2\rangle_J,\nonumber\\[2mm]
&\Gamma(E_{23})&|J_1+1/2,J_2,J_3+1/2,R\rangle_J\nonumber\\
&=&2\left[J_1+R+1/2\right]_q|J_1,J_2,J_3,R-1/2\rangle_J,\nonumber\\[2mm]
&\Gamma(E_{23})&|J_1,J_2+1/2,J_3+1,S\rangle_J\nonumber\\
&=& \frac{\left[J_1+S\right]_q}{2D_4 (J_1-S)[2J_1+1]_q} \nonumber \\
 &&\times |J_1-1/2,J_2+1/2,J_3+1/2,S-1/2\rangle_J \nonumber \\
 &&- \frac{1}{2D_4(J_1-S)\left[ 2J_1  + 1 \right]_q }|J_1+1/2,J_2,J_3+1/2,S-1/2\rangle_J.
\end{eqnarray}
\begin{eqnarray}
\label{E32}
&\Gamma(E_{32})&|J_1,J_2,J_3,M\rangle_J\nonumber\\
&=& \frac{\left[J_1-M\right]_q\left[2J_1+2J_2+1\right]_q}{2\left[ {2J_1  + 1} \right]_q }\nonumber \\
&& \times |J_1-1/2,J_2+1/2,J_3+1/2,M+1/2\rangle_J \nonumber \\
&&  + \frac{\left[ {2J_2 } \right]_q }{2\left[ {2J_1  + 1} \right]_q }|J_1+1/2,J_2,J_3+1/2,M+1/2\rangle_J
 ,\nonumber\\[2mm]
&\Gamma(E_{32})&|J_1-1/2,J_2+1/2,J_3+1/2,P\rangle_J\nonumber\\
&=& 2\left[2J_2\right]_qD_4 (J_1-P-1/2) \nonumber \\
&&\times |J_1,J_2+1/2,J_3+1,P+1/2\rangle_J,\nonumber\\[2mm]
&\Gamma(E_{32})&|J_1+1/2,J_2,J_3+1/2,R\rangle_J\nonumber\\
&=&-2[J_1-R+1/2]_q\left[2J_1+2J_2+1\right]_q D_4 (J_1-R - 1/2) \nonumber \\
 && \times  |J_1,J_2+1/2,J_3+1,R+1/2\rangle_J ,\nonumber\\[2mm]
&\Gamma(E_{32})&|J_1,J_2+1/2,J_3+1,S\rangle_J=0.
\end{eqnarray}

The next step is to investigate the irreduciblity of the representations \eqref{Hi}--\eqref{E32}. It can be shown that the latter are irreducible and called typical if the condition
\begin{equation}
[2J_1+2J_2+1][2J_2]\neq 0
\label{typical}
\end{equation}
holds (cf. \cite{KV03}). When \eqref{typical} is violated, i.e., $[2J_1+2J_2+1][2J_2]=0$, the constructed representations \eqref{Hi}--\eqref{E32} are no longer irreducible but indecomposable. However, in an indecomposbale representation space (module) of $U[gl(2|1)]$, there is always an invariant subspace. Factorizing an indecomposable module $W$ by an invariant subspace $I_k$ gives rise to an irreducible representation in the factor module $W/I_k:= W^{(k)}$. The obtained irreducible reperesentation is called nontypical.
It can be proven that the $U[gl(2|1)_{\bar{0}}]$-submodule $V_1$ always belongs to an invariant subspace (in an indecomposible module), which never contains $V_4$.  \\

There are two classes of nontypical representations corresponding to two ways of the violation of \eqref{typical}:\\
\begin{equation}
[2J_1+2J_2+1]=0 ~~\mbox{but}~~ [2J_2]\neq 0
\label{nontyp-1}
\end{equation}
or
\begin{equation}
[2J_2]=0 ~~ \mbox{but}~~ [2J_1+2J_2+1]\neq 0. \label{nontyp-2}
\end{equation}
We call \eqref{nontyp-1} and \eqref{nontyp-2} nontypical conditions. To save the length of the paper we do not write down matrix elements of nontypical representations but we just explain briefly how they can be derived. The case when both $[2J_1+2J_2+1]=0$ and $[2J_2]=0$ is excluded because it leads to trivial representations in which all the matrix elements of $E_{31}$ and $E_{32}$ are identically equal to zero.\\[4mm]
{\it Class-1 nontypical representations}: These nontypical representations are denoted so in correspondence with the first nontypical condition \eqref{nontyp-1}. The invariant suspace in this case is $I_1=V_1\oplus V_3$. Then, the matrix elements of this class nontypical representations can be found from \eqref{Hi}--\eqref{E32} by applying \eqref{nontyp-1} and replacing all vectors belonging to $I_1$ by 0.\\[4mm]
{\it Class-2 nontypical representations}: These nontypical representations correspond to \eqref{nontyp-2}. The invariant suspace in this case is $I_2=V_1\oplus V_2$. Then, the matrix elements of this class nontypical representations can be found from \eqref{Hi}--\eqref{E32} by applying \eqref{nontyp-2} and replacing all vectors belonging to $I_2$ by 0.\\

Let us emphasized that the constructed typical and nontypical representations cover all finite-dimensional irreducible representations of $U_q[gl(2|1)]$. We skipt the proof of this statement which is a bit long
but it can be done in a similar way as in \cite{Uqgl22-1,Uqgl22-2}. The latter works investigated the quantum superalgebra $U_q[gl(2/2)]$ in another, namely, Gelfand-Tsetlin basis which, however, unlike the CS basis,
gives no direct physics imagination.\\

It can be checked that the generators \eqref{Gamq}, i.e., \eqref{Hi}--\eqref{E32}, continue to
satisfy the commutation relations of $U_q[gl(2|1)]$ if the functions $D_i(N_{a_{12}})$
given in  \eqref{func} are replaced by arbitrary functions of $N_{a_{12}}$, say $F_i (N_{a_{12}})$.
Therefore, the generalized mapping
\begin{eqnarray}
 \left| \psi  \right\rangle & \to & \left| \psi_{gen}  \right\rangle _J=\left\langle {J } \right|\left \{F_4 (N_{a_{12} } )\alpha_{13}^\dag  E_{13}
\alpha_{23}^\dag  E_{23}+ F_3 (N_{a_{12} } )\alpha_{13}^\dag  E_{13\notag }\right.\\&&  +
\left.e^{F_2 (N_{a_{12} } )a_{12}^\dag  \alpha_{23}^\dag  E_{13} } e^{F_1 (N_{a_{12} } )\alpha_{23}^\dag
E_{23} } \right \}e_q^{a_{12}^\dag  E_{12} } \left| \psi  \right\rangle \left| 0 \right\rangle,\label{Fa1}\\[2mm]
A\left| \psi  \right\rangle &\to & \Gamma (A)\left| \psi_{gen}  \right\rangle _J  =
\left\langle {J } \right|\left \{F_4 (N_{a_{12} } )\alpha_{13}^\dag  E_{13} \alpha_{23}^\dag  E_{23}
+ F_3 (N_{a_{12} } )\alpha_{13}^\dag  E_{13}\right. \notag  \\
&& \left. + e^{F_2 (N_{a_{12} } )a_{12}^\dag  \alpha_{23}^\dag  E_{13} }
e^{F_1 (N_{a_{12} } )\alpha_{23}^\dag  E_{23} } \right \}e_q^{a_{12}^\dag  E_{12} }
A\left| \psi  \right\rangle \left| 0 \right\rangle \label{Fa},
\end{eqnarray}
containing \eqref{eq2} and \eqref{eq3} as a special case, also realizes representations of $U_q[gl(2|1)]$,
i.e., we obtain a class of boson-fermion realisations and representations more general than \eqref{Gamq}
and  \eqref{Hi}--\eqref{E32}. (Putting $q=1$, we get the corresponding generalized boson-fermion realisations
and representations of $gl(2|1)$). The matrix elements of the generalized representations can be obtained
from those in a CS basis by replacing $D_i(N_{a_{12}})$ with $F_i(N_{a_{12}})$. Now, the analysis made
above on the representation irreducibility can be applied to obtain, thus, generalized typical and nontypical representations. The typical and nontypical conditions (\ref{typical})--(\ref{nontyp-2}) remain unchanged as
the explicit forms of $D_i(N_{a_{12}})$ were not used in obtaining these conditions. Therefore, the structure
of irreducible (i.e., typical and nontypical) and indecomposable representations remains the same as that of
the representations in a CS basis \eqref{Gamq} and \eqref{Hi}--\eqref{E32}. An open problem here is to find
an explicit physical interpretation which may fix the form of $F_i (N_{a_{12}})$.\\

A special interest is the case when $q$ is a root of unity (undertood to be different from $\pm 1$).
However, we keep it for a later investigation because the irreduciblity and indecomposiblity structure
in this case (having often additional invariant subspaces) is drastically differerent from that
in the case of a generic $q$.
\section{Conclusion}

In this paper, representations ($q$--boson-fermion realisations and finite-dimensional representations) of quantum superalgebra $U_q[gl(2|1)]$, which are both mathematically and physically interesting, have been constructed by the coherent state method playing an important role in mathematics and quantum physics.
With this result, to our knowledge, $U_q[gl(2|1)]$ becomes one of the first, in fact, the biggest so far, quantum superalgebras with boson-fermion realisations and representations constructed explicitly by the CS method. The representation irreducibility is also considered and as a result irreducible representations are classified into typical and nontypical representations. The latter in turn are classified into two sub-classes whose matrix elements can be obtained from those given in \eqref{Hi}--\eqref{E32} by applying the corresponding nontypical conditions \eqref{nontyp-1} and \eqref{nontyp-2} and replacing all the vectors belonging to invariant subspaces by zero. It can be proved that these typical and nontypical representations constitute all irreducible finite-dimensional representations of $U_q[gl(2|1)]$.\\

Furthermore, a general class of representations of $U_q[gl(2|1)]$ and, thus, of $gl(2|1)$ have been obtained. All the analyses on the representation irreducibility made above can be applied to this case leading to generalized typical and nontypical representations of $U_q[gl(2|1)]$ and $gl(2|1)$. \\

We hope that the results obtained will be useful for construction of representations of bigger quantum superalgebras as well as for investigating different fields in mathematics and quantum physics such as non-commutative geometry, current superalgebras, CFT's, integrable systems, string theory, quantum gravity and cosmology, etc. In principle, the present method can be applied to constructing representations in a CS basis of higher rank quantum superalgebras, e.g., $U_q[gl(2|2)]$, \cite{Uqgl22-1,Uqgl22-2} but
the latter will be certainly technically cumbersome, therefore, the method needs a further development.\\[4mm]
\textbf{Acknowledgements}:
One of the authors (N.A.K.) would like to thank H. Orland  for kind hospitality at IPhT, CEA, Saclay, and O. Babelon for kind hospitality at LPTHE, Universite Pierre et Marie Curie, Paris. He also acknowledges the warm hospitality of L. Alvarez-Gaume and J. Ellis at the Theory Unit, CERN, Geneva.
%Discussions  and participation of Nguyen Thi Hong Van in
%the first stage of these investigations are acknowledged.
This work is partially supported by the Vietnamese Academy of Science and Technology (VAST) in the framwork of the VAST-CNRS collaboration program in high energy physics and by Vietnam's National Foundation for Science and Technology Development (NAFOSTED) under the grant 103.02.112.09.

\end{document}